\documentclass[twocolumn]{elsart}

\usepackage{natbib}

 \usepackage{graphicx}

\usepackage{amssymb}

\begin{document}

\begin{frontmatter}



\title{Lepton flavor violating process in a supersymmetric
grand unified theory with right-handed neutrino
\thanksref{talk}
}
\thanks[talk]{Talk given by K.~Okumura at the 3rd International Workshop
 on Neutrino Factories based on Muon Storage Rings (NuFACT'01
Workshop) Tsukuba, Japan, May 24-30, 2001}

\author[label1]{Seungwon Baek}
\author[label2]{Toru Goto}
\author[label2,label3]{Yasuhiro Okada}
\author[label4]{Ken-ichi Okumura}

\address[label1]{Department of Physics, National Taiwan University,
Taipei 106, Taiwan}
\address[label2]{Theory Group, KEK, Tsukuba, Ibaraki, 305-0801, Japan}
\address[label3]{Department of Particle and Nuclear Physics, The
Graduate University of Advanced Studies, Tsukuba, Ibaraki, 305-0801, Japan}
\address[label4]{Institute for Cosmic Ray Research, University of
Tokyo, Kashiwanoha 5-1-5, Kashiwa, 277-8582, Japan}

\begin{abstract}
Motivated from the recent results of neutrino oscillation experiment,
we investigated lepton flavor violating (LFV) processes
 in a SU(5) supersymmetric grand unified theory with right-handed neutrino.
The current experimental upper bound for $\mu \to e\gamma$ process
gives already some constraint on the model. 
Correlation between $\mu \to e \gamma$ and
 the SUSY contribution to the muon anomalous magnetic moment
 is also discussed.
Future LFV experiments will give considerable impacts
 on this type of SUSY GUT models equipped with seesaw neutrino mass
 generation.
\end{abstract}


\end{frontmatter}


Supersymmetric standard model (SUSY SM) is a promising candidate
 for physics beyond the SM.
Supersymmetry resolves elegantly the hierarchy problem of the SM.
Furthermore, gauge coupling unification is realized
 if we assume grand unification with the SU(5) gauge group or its
extensions.
On the other hand, the seesaw mechanism naturally explain
 the smallness of neutrino mass which is suggested from
 the recent neutrino oscillation experiments. 
 It introduces heavy Majorana right-handed
 neutrinos and their Yukawa interaction to lepton doublets.
In the SUSY models, it is known that
 a radiative correction  to the slepton
 mass matrix from this Yukawa interaction induces
 lepton flavor violating (LFV) processes
 such as $\mu \to e\gamma$ which is forbidden in the SM. 

In this paper, we show the results of our numerical
analysis for the LFV processes in a SU(5)
 supersymmetric grand unified theory (SUSY GUT) 
 with right-handed neutrino
\footnote{This talk is based on the work in Ref.\cite{01BaGoOkOk}.}
.
We calculated branching ratios of $\mu \to e \gamma$,
 $\tau \to \mu \gamma$, and $\tau \to e\gamma$,
 and the SUSY contribution to the muon anomalous magnetic moment,
 for which recently BNL E821 experiment reported $2.6\sigma$
 deviation from the SM theoretical calculation \cite{g-2exp}.
In our analysis, we introduced higher dimensional operators
 above the GUT scale to accommodate realistic mass relations 
 between the down-type quarks and the charged leptons.
Consequently, new degrees of freedom appear at the GUT scale when we
 embed the fields of minimal SUSY SM (MSSM)
 in the SU(5) multiplets as
${\bf 10}^i[Q^i,(V_U)^i_j\overline{U}^j,(V_E)^i_j\overline{E}^j]$,
${\bf \overline{5}}^i[(V_D)^i_j\overline{D}^j,L^i]$
 where $V_U$, $V_E$, and $V_D$ are unitary matrices
 defined in the basis where the  mass matrices of
 down-type quark and charged lepton are diagonal.
In the actual analysis, we considered the effects of two mixing
 angles $\theta_E$ in $V_E$ and $\theta_D$ in $V_D$ which describe the
 rotation between the first and the second generations. 
We numerically solved renormalization group equations from the 
 Planck scale to the EW scale.
 For the initial condition, we assumed
 universal scalar and gaugino masses $m_0$, $M_0$
 and SUSY breaking scalar couplings
 are taken to be proportional to the superpotential
 with a universal coefficient
 $m_0 A_0$.

\begin{figure} 
\begin{center}
\makebox[0em]{
\includegraphics*[height=7cm]{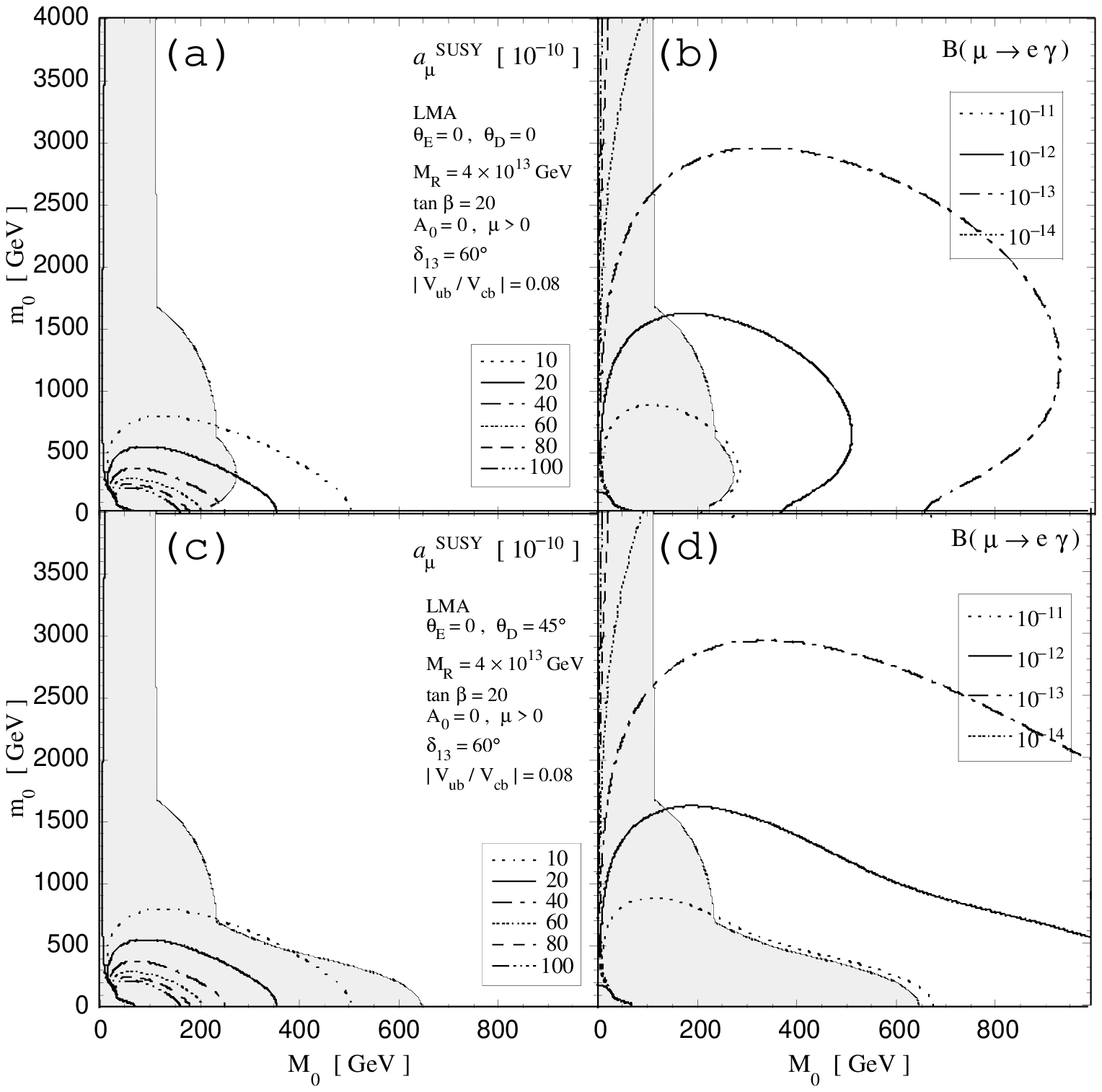}
}
\end{center}
\caption{Contour plots of $a^{SUSY}_{\mu}$ and $B(\mu \to e\gamma)$
 in the $m_0$-$M_0$ plane.}
\label{fig:fig1}
\end{figure}
Fig.\ref{fig:fig1} (a) and (b) shows the contour plots of the SUSY contribution
 to the muon anomalous magnetic moment $a^{SUSY}_{\mu}$ and
 $B(\mu \to e\gamma)$ in the $m_0$-$M_0$ plane.
We take $\tan\beta=20$, $\mu>0$, $A_0=0$,  $\theta_D=\theta_E=0$,
 and $M_R=4.0\times 10^{13}$ GeV where $M_R$ is the Majorana mass
scale of the right-handed neutrinos.
We assumed the MSW large mixing angle solution for the
 solar neutrino anomaly.
The shaded region is excluded by the direct SUSY searches
 at LEP and Tevatron, the Higgs boson search at LEP,
 the result of $B(b \to s \gamma)$ at CLEO, 
 and $B(\mu \to e\gamma)$.
The current experimental upper bound, $B(\mu \to
e\gamma)<1.2\times10^{-11}$ 
 already gives some constraint on the model.
The favored region of BNL E821 experiment
 which corresponds to $20\times 10^{-10}
 \lesssim a^{SUSY}_{\mu} \lesssim 60 \times 10^{-10}$
 still survives within these constraints.
An experimental improvement by order of magnitude
 could change the situation.  
If we lower $M_R$,
 the $\mu \to e\gamma$ constraint becomes
 weak while $a_{\mu}^{SUSY}$ is intact,
 however, in such a case,
 ${y_{\nu}}_3$ becomes much smaller than $y_t$ and
 further GUT unification scenario such as SO(10) SUSY GUT
 becomes difficult.
If we lower $\tan\beta$, not only $B(\mu \to e\gamma)$ but also
 $a_{\mu}^{SUSY}$ decreases and
 the constraint from the Higgs boson search becomes stronger. 
With the same parameter set,
 $B(\tau \to \mu \gamma)$ is $\lesssim 10^{-8}$ and
 $B(\tau \to e \gamma)$ is far below the current experimental bound.
$\mu$-$e$ conversion in atomic nuclei is an another important
 LFV process.
When the process is dominated by the photon-penguin diagrams
 as in our case, it is known 
 that the conversion rate is related to $\mu \to e\gamma$ as
 $B(\mu^-\,N\to e^-\,N)/B(\mu^+\to e^+\,\gamma)
  \approx
  B(A,Z)/428$
where $B(A,Z)\approx1.1$ for
$^{27}$Al, $B(A,Z)\approx1.8$ for $^{48}$Ti and $B(A,Z)\approx1.25$ for
$^{208}$Pb.

\begin{figure} 
\begin{center}
\includegraphics*[height=4.8cm]{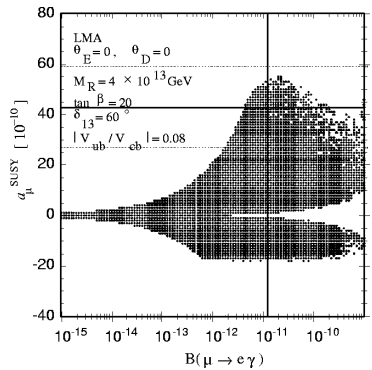}
\end{center}
\caption{Correlation between $a^{SUSY}_{\mu}$ and $B(\mu \to e\gamma)$.}
\label{fig:fig2}
\end{figure}

Fig.\ref{fig:fig1} (c) and (d) shows the same plots for the
 case $\theta_D=\pi/4$.
Other parameters are same as (a) and (b). 
Similar plots can be drawn for $\theta_E=\pi/4$.
Inclusion of the effects of higher dimensional operators
 generally increases $B(\mu \to e\gamma)$ especially in the small $m_0$
region.
The parameter region where the BNL E821 experiment
 favors is excluded in this figure.
It is found that this contribution does not disappear even if we
 lower $M_R$ because it mainly comes from the flavor
 mixing in LR mixing mass which originates
 in the matching condition of SUSY breaking scalar couplings
 at the GUT scale rather than the radiative corrections
 from the neutrino Yukawa coupling.
$B(\tau \to \mu \gamma)$
 is not affected by $\theta_D$ and $\theta_E$ and
$B(\tau \to e \gamma)$ remains negligible.
Fig.\ref{fig:fig2} shows a correlation between the $a_{\mu}^{SUSY}$ and the
 $B(\mu \to e\gamma)$ where we searched SUSY parameters as $m_0, M_0 <
3$ TeV,
 $|A_0|<5$. We also took both sign of $\mu$ parameter.
 Other parameters are same as Fig.\ref{fig:fig1} (a) and (b).
The dotted lines show the deviation
 reported by the BNL E821 experiment
 and the vertical line indicates the current
experimental upper bound for $B(\mu \to e\gamma)$.
It is clearly seen that future LFV experiments will
 give considerable impacts on this type of SUSY GUT models combined with
 the further improvement of our understanding
 on the muon anomalous magnetic moment.

In summary, LFV processes
 provide us a unique opportunity testing the physics
 at the GUT scale and the Majorana mass scale which 
 is relevant with the origin of neutrino masses and mixing
 and future experimental progress is expected to bring some insights
 on it.

\end{document}